# Multifunctional all-dielectric nano-optical systems using collective multipole Mie resonances: Towards on-chip integrated nanophotonics

Swarnabha Chattaraj [1] and Anupam Madhukar [2,3, *]

[1] Ming Hsieh Department of Electrical Engineering, University of Southern California, Los Angeles, CA 90089, USA

[2] Department of Physics and Astronomy, University of Southern California, Los Angeles, CA 90089, USA

[3] Mork Family Department of Chemical Engineering and Materials Science, University of Southern California, Los Angeles, CA 90089, USA

[*] Corresponding Author Email: madhukar@usc.edu

## ABSTRACT

We present an analysis of the optical response of a class of on-chip integrated nano-photonic systems comprising all-dielectric building block based multifunctional light manipulating units (LMU) integrated with quantum dot (QD) light sources. The multiple functions (such as focusing excitation light, QD emission rate enhancement, photon guidance, and lossless propagation) are simultaneously realized using the collective Mie resonances of dipole and higher order multipole modes of the dielectric building blocks (DBBs) constituting a single structural unit, the LMU. Using analytical formulation based on Mie theory we demonstrate enhancement of the excitation light simultaneously with the guiding and propagation of the emitted light from a QD emitter integrated with the DBB based LMU. The QD-DBB integrated structures can serve as the basic element for building nano-optical active circuits for optical information processing in both classical and quantum realms.

# I. INTRODUCTION

Collective optical response of structures made of subwavelength-sized high refractive index dielectric building blocks (DBBs) are being increasingly examined for their potential use in on-chip integrated nano photonic systems for manipulation of light. The interaction of light with such dielectric structures lies predominantly in the collective magnetic multipole resonances of the interacting array of the sub-wavelength size constituent DBBs which are fundamentally distinct from those of the individual BBs. Thus by controlling the geometry, separations, and optical properties of the DBBs one can access considerable richness in control over the response of the system that simply cannot be realized in a continuous medium. Our conception of such on-chip integrated nano-optical systems is schematically illustrated in Fig.1. The blue pillars depict the DBBs and the pyramidal structure represents a single quantum dot (SQD) as a light source or detector. We note that spatially regular arrays of single quantum dot based single photon emitters, well suited for integration with DBB based LMUs, have been realized [1, 2, 3]. The theoretical studies reported here are motivated by such experimental developments and provide a framework for the co-design of the light emitting and manipulating components of the integrated systems to be experimentally realized.

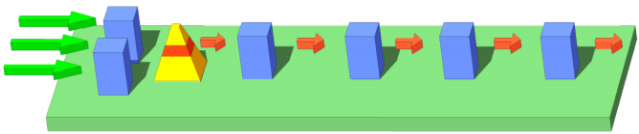

**Figure 1** A schematic of an on-chip integrated nanophotonic system comprising multifunctional-structure made of dielectric building blocks (blue blocks) and a quantum dot light emitter or detector.

We note that the ~100 nanometer size scale of the high index dielectric BBs also offers a way of reducing the footprint of light manipulating passive elements compared to other approaches such as photonic crystal cavity [4, 5]. Although typically in dielectric systems the achievable confinement volumes of the electromagnetic modes are not as small as achievable in plasmonic systems [6, 7], the absence of intrinsic conductive loss makes large area on-chip integrated optical circuits possible. The use of nanoscale (~100nm) dielectric building blocks allows subdiffraction manipulation of light and also realization of negative permittivity and negative permeability medium at optical frequency [8, 9]. Collective resonances in DBB systems outside the negative index regime such as Fano type resonance [10] in finite DBB clusters, and collective modes in DBB arrays with period of the order of $\lambda/2$ are also of interest. In fact, lossless and long-range waveguiding with subdiffraction footprint in coupled array of high index dielectric building blocks has been a subject of study recently [11, 12, 13, 14]. Subwavelength beam steering can be achieved by exploiting the symmetry of different multipole modes in the DBBs [15]. Similar principles have been used to demonstrate nanoantenna structures that can boost the directivity of the radiation from a dipole like source towards a desired direction [16, 17, 18]. Furthermore, such array of DBBs with predominantly dipole mode excitation has been shown to generate Purcell enhancement of the order of hundred for a dipole like emitter [16,19]. However, the common approach of utilizing the dominant electric and magnetic dipole modes in the DBBs to individually implement the different functions does not take advantage of the full potential offered by the higher order multipole modes of the DBBs. Thus to complement the aforementioned studies examining individual functions using only the dipole modes of the DBB arrays, we focus on exploring the use also of the higher order multipole modes in the DBBs to simultaneously implement multiple functions—potentially an important consideration for all-optical integrated photonic system design.

To this end, we present exploration of the performance of on-chip integrated quantum dot light emitter and subwavelength size DBB resonator based nanophotonic systems exemplified by Fig. 1 that exploit the collective resonances of the higher order magnetic and electric multipole modes to provide *simultaneously* functions of light focusing (green arrows) at the excitation (pumping) wavelength of the quantum dot light source (shown as the pyramid, [1]) and guidance and propagation of the light at the emission wavelength (red arrows). Such integrated nanophotonic systems comprising light sources, light manipulation components (such as lens, waveguide, beam splitter, etc.) and detectors that perform required multiple functions at different desired optical wavelengths efficiently will potentially have transformative impact on a wide variety of photonic technologies for applications ranging from information communication, computing, biological imaging, environmental sensing, biomedical platforms, to nanomedicine. In the past decade much rapid progress has been made on both the development of individual components [20] and their hybrid integration on increasingly smaller foot-print platforms [20-23]. For instance, single quantum dots (SQDs) as single photon emitters have been established [24] although efforts to move these towards viable technologies [24] demand overcoming limitations of severe spectral emission nonuniformity while also placing the SQDs in spatially ordered arrays that are readily integrable [2, 3, 25, 26,]. The progress towards *on-chip* integration of various components in a photonic system has been slow and limited to hundreds of micron footprint largely owing to the dimensions of the individual passive elements such as waveguides and beam splitters, typically lithographically curved out of a continuous media [27] or implemented based on photonic crystal cavities [5]. Integration of spatially ordered and spectrally uniform SQDs with the DBB based multifunctional optical systems to realize on-chip integrated nanophotonic systems will bring out new physics and device applications in both classical and quantum optical domains.

We examine this new paradigm for on-chip integrated multifunctional nanophotonic systems [28] by modelling the DBBs as spherical dielectric nanoparticles (hereafter denoted as DNPs) as this enables analytical modeling that, in turn, helps gain clearer physical insight. Thus we focus on light response in a prototypical integrated QD-DNP unit in a planar architecture as depicted schematically in Fig. 2(a). The quantum dot exciting green light is focused at the quantum dot location using a particular multipole (here magnetic octupole) and the light emitted by the quantum dot is guided by another multipole (here magnetic and electric dipoles) of the constituent BBs. Units such as in Fig. 2(a), we anticipate, will serve as building blocks of more complex hierarchical optical networks for system level functionalities. In these initial studies the blue and yellow DNPs are chosen to be identical in material and size to highlight that the multifunctional behavior is achieved using different magnetic and electric multipole modes of the same DNPs. We emphasize that though the spherical shape chosen in these studies permits analytical mathematical description of light scattering, a variety of shapes, including mixed shapes, can be readily examined via numerical analyses for designs optimized for possible functional architectures in planar geometries such as the structure shown in Fig.1 as well as other planar and non-planar geometries. Panels (b) and (c) of Fig.2 show the calculated electric field distributions corresponding to these functions at the two wavelengths. The details of the analysis underlying the study of these functions based on the collective multipole modes and the specific case results shown in Fig.2 constitute the body of this paper.

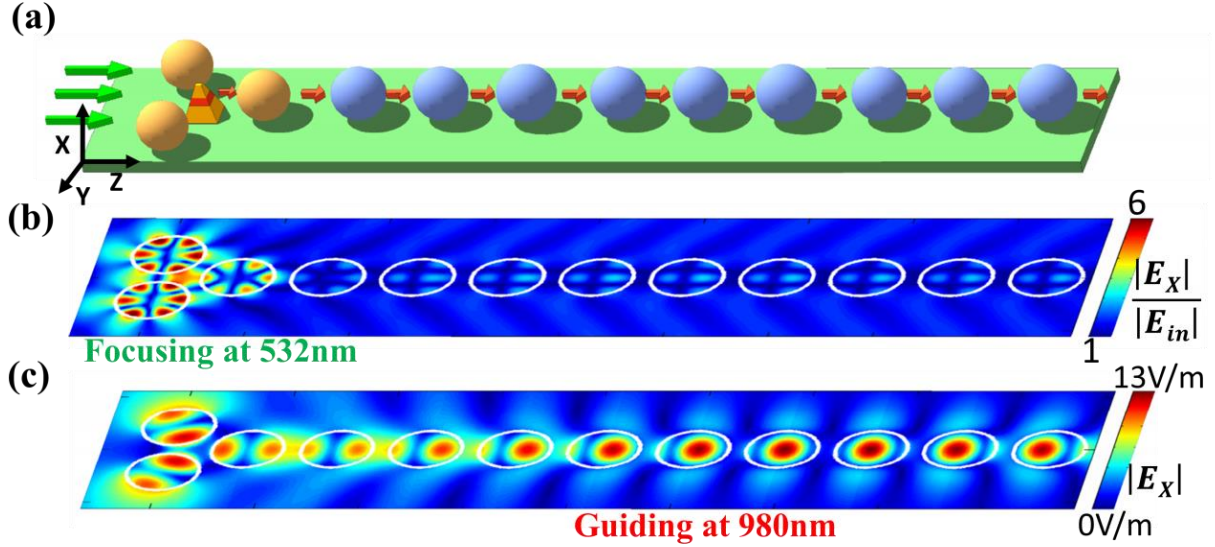

**Figure 2. (a)** Schematic of the analyzed unit of a SQD integrated with spherical dielectric nanoparticle (DNP) based multifunctional optical system. The calculated electric field (normalized to the incident field) distribution in the YZ plane (horizontal plane passing through the center of the DNPs) are shown in panel (b) for the focusing of the QD exciting light at 532nm (green arrows in panel a) using the collective magnetic octupole mode of the DNPs and (c) for the guiding of the QD emitted light at 980nm (red arrows in panel a) using the collective magnetic and electric dipole modes of the DNPs. The electric field distribution is normalized to 1 Debye of QD transition dipole strength.

## II. FORMULATION OF THE OPTICAL RESPONSE:

## MULTIPOLE EXPANSION

The optical response of the interacting assembly of spherical DNPs is calculated using analytical approach based on Mie theory [29, 30] that provides fast and accurate computation compared to any conventional FEM or FDTD based method and also provides physical insight of multipole scattering facilitating device design. In this approach, the solution of the scattered wave by the DNP is obtained in the form of the spherical vector harmonics that not only form a complete basis for electromagnetic fields but are natural eigenfunctions of Maxwell's wave equations. Hence they form independent modes also known as multipole modes in a spherically symmetric system. The spherical vector harmonics can be expressed in terms of the scalar harmonics as,

$$\bar{m}_{q,p} = \bar{\nabla} \times (\bar{r} \psi_{q,p}) \tag{1}$$

and,

$$\bar{n}_{q,p} = \bar{\nabla} \times (\bar{\nabla} \times (\bar{r}\psi_{q,p})) \qquad (2)$$

where $\psi_{q,p} = \sqrt{q(q+1)} z_q(\beta r) Y_{q,p}(\theta,\phi)$, $z_q(\beta r)$ being the appropriate spherical Bessel function. The $\bar{m}_{q,p}$ and $\bar{n}_{q,p}$ complementarily describe the electric and magnetic field components of a propagating spherical wave for the two types of multipole modes, named TE$_R$ (Transverse electric field to the radial direction) and TM$_R$ (Transverse magnetic field to the radial direction) modes, also known as magnetic and electric multipole modes respectively. These two types of multipole modes can be mathematically expressed as,

$$\bar{E}_{TE_{q,p}} = \bar{m}_{q,p}, \bar{H}_{TE_{q,p}} = \frac{1}{i\eta} \bar{n}_{q,p} \qquad (3)$$

and

$$\bar{E}_{TM_{q,p}} = \bar{n}_{q,p}, \bar{H}_{TM_{q,p}} = \frac{1}{i\eta} \bar{m}_{q,p} \qquad (4)$$

where $\eta$ represents the impedance of the surrounding medium.

For an $N$ particle assembly with the particle center positions denoted by the vectors $\{\bar{r}_i\}_{i=1}^N$, the expansion of the scattered electric and magnetic fields under the multipole mode basis of individual DNPs takes the form [30],

$$\bar{E}(\bar{r}) = \sum_{i=1}^{N} \sum_{n=1}^{n_{max}} \sum_{m=-n}^{n} (a_{n,m}^{(i)} \bar{E}_{TE_{n,m}}(\bar{r}-\bar{r}_i) + b_{n,m}^{(i)} \bar{E}_{TM_{n,m}}(\bar{r}-\bar{r}_i)) \qquad (5)$$

$$\bar{H}(\bar{r}) = \sum_{i=1}^{N} \sum_{n=1}^{n_{max}} \sum_{m=-n}^{n} (a_{n,m}^{(i)} \bar{H}_{TE_{n,m}}(\bar{r}-\bar{r}_i) + b_{n,m}^{(i)} \bar{H}_{TM_{n,m}}(\bar{r}-\bar{r}_i)) \qquad (6)$$

where $i$ denotes the particle index in the nanoparticle assembly. We note that the Bessel function used in the expression of the scattered wave here is the spherical Hankel function of type 1, representing the radially outward waves for $e^{-i\omega t}$ time evolution. Under this expansion, the scattered wave is represented by the set of coefficients $\{a_{n,m}^{(i)}, b_{n,m}^{(i)} | i=1,2...N; n=1,2,...n_{max}; m=-n,-(n-1),...n\}$. For finiteness, the series needs to be terminated at an $n = n_{max}$. The value of $n_{max}$ need to be set to achieve desired accuracy for particular cases.

All the TE and TM modes are normalized with respect to the scattered power, where the power is given by,

$$P_{Rad} = \frac{|a_{q,p}^2|(or|b_{q,p}^2|)}{2\beta^2 \eta}, \qquad (7)$$

so that, just comparing the mode coefficients gives us a sense of the power radiated by a particular mode, which will be useful in optimization of the efficiency of the DNP structures in the subsequent sections.

The next step in the mathematical formulation of the problem is to obtain the spectral response of these multipole mode coefficients under the influence of an arbitrary source. As the source itself is expanded in the basis of the multipole modes, the DNP's respond to the source field with TE and TM mode susceptibilities denoted by $\Gamma_{TEn,m}$ and $\Delta_{TMn,m}$, respectively [30]. For spherical nanoparticles, the $\Gamma_{TEn,m}$ and $\Delta_{TMn,m}$ can be derived analytically as,

$$\Gamma_{TE_{n,m}} = \frac{-n_i \hat{j}_n(\beta R)\hat{j}'_n(\beta_d R) + \hat{j}'_n(\beta R)\hat{j}_n(\beta_d R)}{n_i \hat{h}^{(1)}_n(\beta R)\hat{j}'_n(\beta_d R) - \hat{h}'^{(1)}_n(\beta R)\hat{j}_n(\beta_d R)} \tag{8}$$

$$\Delta_{TM_{n,m}} = \frac{-n_i \hat{j}'_n(\beta R)\hat{j}_n(\beta_d R) + \hat{j}_n(\beta R)\hat{j}'_n(\beta_d R)}{n_i \hat{h}'^{(1)}_n(\beta R)\hat{j}_n(\beta_d R) - \hat{h}^{(1)}_n(\beta R)\hat{j}'_n(\beta_d R)} \tag{9}$$

For plane wave incidence on a single DNP, only the only the +1 and -1 angular momentum modes (represented by the index m of the multipole mode) are excited corresponding to the right circular and left circular components of the incident wave (shown in Fig.3(a)). As depicted in Fig.3(b), for a linearly polarized wave these two components are excited equally, resulting in a transverse dipole mode with no net angular momentum, which is identical to TE$_{1,0}$ excitation in a rotated frame with Y axis as the axis of the spherical coordinate. For electric dipole and other higher order multipoles, similar effects can be observed where the angular momentum of the incident photon gets transferred to the angular momentum of the multipole mode of the DNP. Figure 3(c) shows the spectrum of the scattering cross section different magnetic and electric multipole modes of a single DNP excited by an incident plane wave (angular momentum $m = \pm 1$). The DNP's radius ($R$) is chosen to 171.3nm and refractive index ($n_i$) to 2.7 which corresponds to TiO$_2$. Note the narrow spectral width of the magnetic octupole mode near 522nm but the relatively broader spectral width of the magnetic dipole near 980nm - features that we exploit in the examination of the multifunctional structure. The tuning of the spectral distribution of the magnetic and electric resonances in a DNP [31-34] is exemplified in Fig. 3(d) where the dependence of the TE$_{1,1}$ (magnetic dipole) and the TE$_{3,1}$ (magnetic octupole) mode peak wavelengths (normalized to the radius of the DNP) are plotted against the refractive index $n_i$ for the range $1.5 < n_i < 4$ encompassing most of the dielectric media, including Si ($n_i \sim 4$). Through judicious choice of such characteristics of the individual DNPs, the response of a collection of *interacting* optical resonators can be designed to give novel functionalities through control of the relative spatial locations of the DNPs, as has been anticipated quite some time back [35].

We find that, for a specific wavelength excitation, the multipole mode coefficients drop exponentially with the mode order $n$ as seen in Fig.3(e) which plots the $a_{n,1}$ coefficients on the log scale upto $n = 10$ for a plane wave scattering at 980nm by a DNP of radius 171.3nm and refractive index 2.7. The modes corresponding to $n \geq 5$ contribute only $\sim 10^{-15}$ of the energy. We thus set $n_{max} = 4$ in our analysis to reach sufficient accuracy. This choice results in 48 different multipole modes centered on each DNP in the assembly.

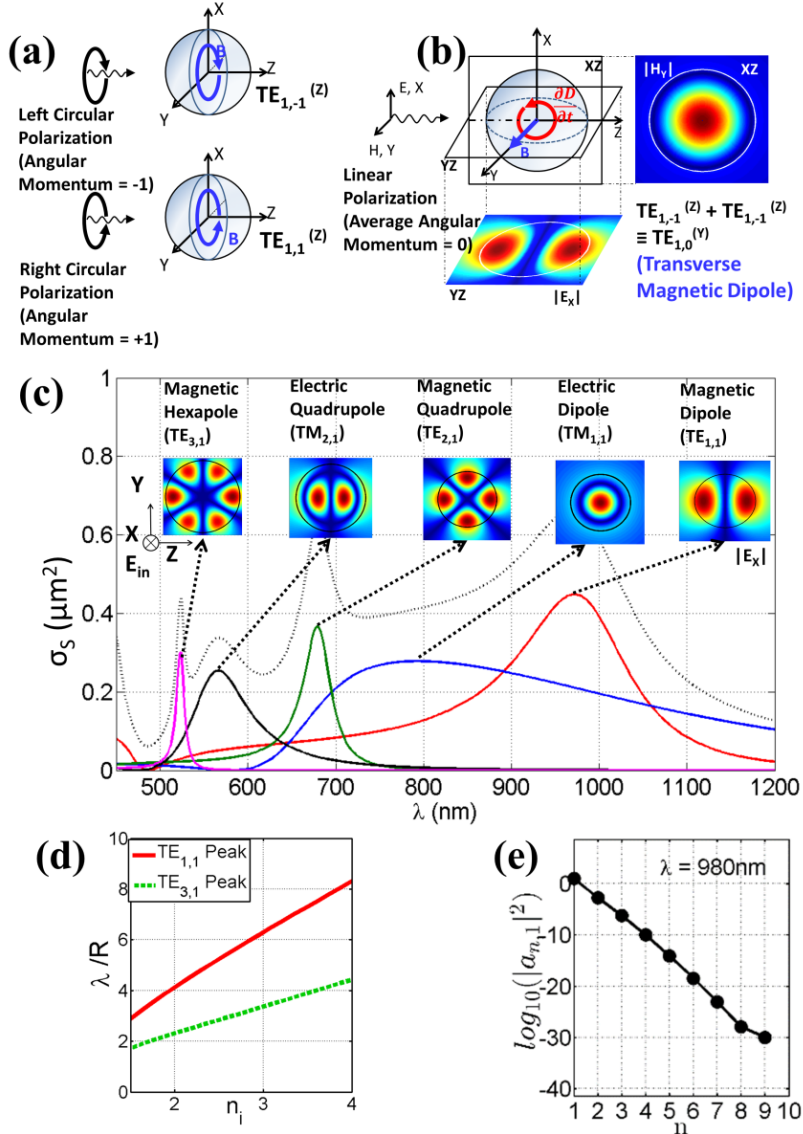

Figure 3.(a) Schematic showing the $TE_{1,-1}$ and $TE_{1,1}$ modes excited by the left and light circular polarized light (b)linear polarized light resulting in transverse magnetic dipole mode. The two color plots represent the E field distribution on the YZ plane and the H field distribution on the XZ plane passing through the center of the particle. **(c)** Spectral dependence of the scattering cross section ($\sigma_S$) of various electric and magnetic multipole modes in a single dielectric nanoparticle of radius $R$=171.3 nm and $n_i$ =2.7 (mimicking TiO$_2$). Insets show the electric field distributions of the multipoles as marked. **(d)** The $TE_{1,1}$ and $TE_{3,1}$ mode peak wavelengths normalized to the radius of the DNP are plotted against the refractive index. (e) Log scale plot of $TE_{n,1}$ mode excitations versus n for plane wave scattering by a DNP of $R$=171.3nm and $n_i = 2.7$ at 980nm.

The interactions between the multipole modes centered on different DNPs are denoted as $\zeta^{i,j}_{q,p,n,m}$ for the $TE^{(j)}_{n,m}$ mode to $TE^{(i)}_{q,p}$ mode as well as $TM^{(j)}_{n,m}$ mode to $TM^{(i)}_{q,p}$ mode ('like' modes) interactions, and $\tau^{i,j}_{q,p,n,m}$ for the $TE^{(j)}_{n,m}$ mode to $TM^{(i)}_{q,p}$ mode as well as $TM^{(j)}_{n,m}$ mode to $TE^{(i)}_{q,p}$ mode ('unlike' modes) interactions. These mode-mode interaction terms can be derived analytically as closed form expressions using the vector spherical harmonic translation theorem [30]. The interacting multi-particle state equation can then be described as,

$$a_{q,p}^{(i)} = \Gamma_{TE_{q,p}} a_{q,p}^{o(i)} + \Gamma_{TE_{q,p}} \sum_{j=1}^{N} \sum_{n=1}^{n_{max}} \sum_{m=-n}^{n} (\zeta_{q,p,n,m}^{i,j} a_{n,m}^{(j)} + \tau_{q,p,n,m}^{i,j} b_{n,m}^{(j)}) \quad (10)$$

$$b_{q,p}^{(i)} = \Delta_{TM_{q,p}} b_{q,p}^{o(i)} + \Delta_{TM_{q,p}} \sum_{j=1}^{N} \sum_{n=1}^{n_{max}} \sum_{m=-n}^{n} (\zeta_{q,p,n,m}^{i,j} b_{n,m}^{(j)} + \tau_{q,p,n,m}^{i,j} a_{n,m}^{(j)}) \quad (11)$$

where $a_{q,p}^{o(i)}$ and $b_{q,p}^{o(i)}$ represents the coefficients corresponding to the input wave expansion, expanded in the basis of the vector spherical harmonics with the spherical Bessel function of type 1 (Bessel J). The final form of the equation (10) and (11) results in a 48$N$ dimensional matrix equation which is solved using matrix inversion. The complete description of the scattered field by the DNP assembly is thus reduced to a vector (comprising the *a* and *b* coefficients) of dimension 48$N$, allowing fast optimization of large structures with minimal computational demands.

This analytic method is applied to our multifunctional nanoparticle assembly (designed to interact with the QD light emitter) to obtain the EM fields and extract useful physical insight by analyzing the amplitudes and phases of the multipole modes as described in the next section.

### III. LIGHT FOCUSING AND PROPAGATION

Our aim, as noted in Sec.I, is to examine the multiple functionality required of *on-chip* integrated passive optical elements co-designed and integrated with optical emitters, and eventually detectors. For such systems, the multiple functions that are required encompass enhancing the emission of the source via focusing of the excitation light, improving emission rate and directivity, and on-chip propagation (and improving absorption by the detectors). Most importantly, simultaneous functions at different wavelengths i.e. focusing at the excitation wavelength of an optically excited source such as a QD and guiding and propagation at the emission wavelength is a highly desired configuration for system design. We therefore present findings for focusing of the QD excitation light at 532nm using the magnetic octupole mode and guiding and propagation of the QD emitted light at 980nm using the magnetic and electric dipoles as depicted in Fig. 2. This choice is, in part, motivated by our on-going experimental efforts on the fabrication of such *on-chip* integrated systems involving InGaAs/GaAs nanotemplate-directed single quantum dot photon emitter arrays [1, 2, 3].

### A. Focusing at Excitation Wavelength

In this subsection we present the analysis underlying the finding of the electric field distribution associated with the 532nm light employed for exciting the quantum dot embedded in the multifunctional structure, shown in Sec.I, Fig. 2(b).

In our design the magnetic octupole mode (Fig. 4(a)) is exploited to achieve focusing of 532nm light with the simple *three-particle planar geometry* shown in the inset of Fig.4 (b) with the particles numbered from 1 to 3 for easy reference. The triangular geometry is designed to specifically take advantage of the collective resonance of the triangular symmetric octupole mode, and other arrangements may be considered for achieving the same function via some other multipole mode. The three DNPs, each of radius 171.3nm and refractive index 2.7, are arranged as an equilateral triangle with side length $p_2$. The value of $p_2$ is optimized to get maximum focusing at wavelength 532nm - the frequency doubled NdYAG laser line typically used often for exciting quantum dots that serve as light emitters.

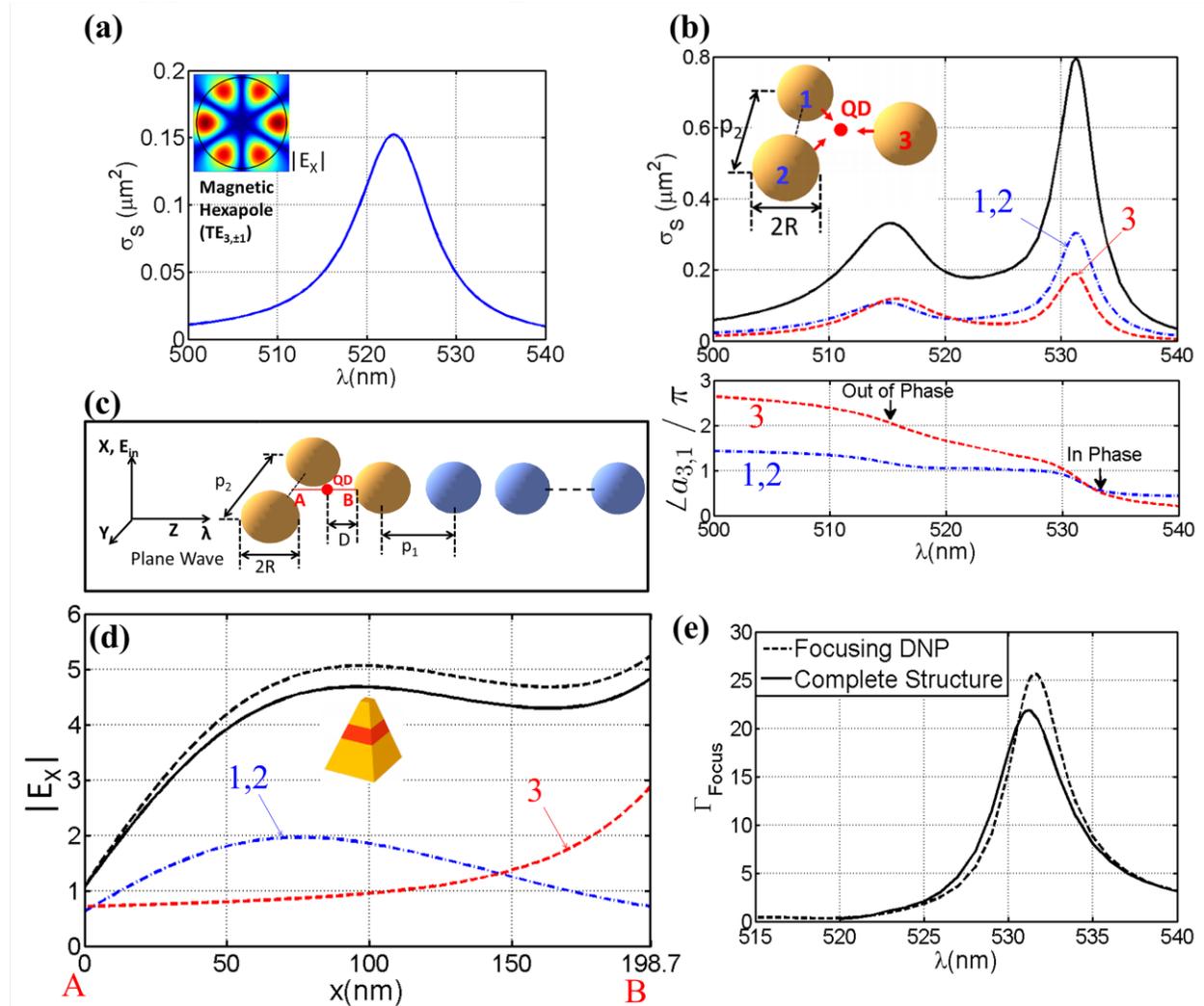

**Figure 4. (a)** The $TE_{3,1}$ (magnetic octupole) mode scattering cross section resonance peak for a single DNP of radius $R$=171.3nm, and $n_i = 2.7$. The inset shows the mode symmetry of the electric field ($|E_X|$); **(b)** The $TE_{3,1}$ mode scattering cross section spectrum of the 3-sphere focusing assembly (upper panel) and the phase of individual DNP resonances (lower panel). Here $p_1$=429nm, $p_2$=427.2nm. The blue dotted curve represents either of DNP 1 and DNP 2, whereas the red dotted curve represents DNP 3. The particle numbers are defined in the geometry of the assembly shown in the inset; **(c)** The geometry of the complete device structure. AB represents a line along the X axis; **(d)** The electric field amplitude (normalized to the incident electric field) at $\lambda$ = 532nm along the line A-B: blue dotted line- contribution from either of DNP 1 or 2; red dotted line from DNP3; broken black line- combined electric field for the three-nanoparticle case; whereas the solid black line- combined electric field for the complete structure; **(e)** The spectral spread of the focal spot electric field intensity enhancement for the three-sphere lens structure (broken line). For comparison, the solid line shows the equivalent spectrum for the complete structure.

As discussed in the preceding section, the linearly polarized incident light excites the +1 and -1 angular momentum modes ($TE_{3,1}$ and $TE_{3,-1}$ modes in this case) in equal proportion, resulting in a single resonance peak shown in Fig.4(a) with the resulting field distribution shown in the inset of Fig.4(a). The spectral response of the scattering cross-section of this three-particle assembly is shown (solid line) in Fig. 4(b). Note that the individual DNP magnetic octupole modes at 522nm (Fig. 4(a)) combine to give two peaks (at ~515nm and ~532nm) owing to the mode-mode interaction between the DNPs. The contributions of the individual particles in the assembly, shown as broken lines, reveal the symmetry-equivalent nature of DNPs 1 and 2, but not 3, with respect to the direction of incidence of the plane wave. The phase spectrum of the magnetic octupole mode of the individual DNPs, plotted in the lower panel of Fig. 4(b), reveals that the peaks at 532nm and 515nm are, respectively, the in-phase and out-of-phase normal modes of the focusing assembly. The constructive interference of the in-phase normal mode forms the basis of the focusing function at the location of the QD in our proposed system.

In Fig.4(d) we show the in-phase mode (532nm) total electric field amplitude profile (dotted black line) along with contributions from the individual nanoparticles (blue broken line: DNP 1 and 2, red broken line: DNP 3) along the axial line A-B in Fig.4(c). An enhancement of the electric field in a large focal spot is realized with a maximum electric field amplitude enhancement of ~5 at a distance of 98.5nm from the surface of the DNP 3. The enhancement remains essentially flat (within 95% of the maximum) for a broad area ~ 50nm scale which is sufficiently large compared to the 10 nm scale confinement of the electronic state in a typical quantum dot. This ensures uniform excitation of the QD in the focal spot as well as grants fabrication tolerance for real device implementations.

The focusing at 532nm is dominated by the local collective octupole oscillation in the triangular DNP assembly (yellow DNPs) and is minimally affected by the blue DNP array. The solid black line in figure 4(d) shows the E field distribution calculated for the complete structure (blue and yellow DNP), showing a minor deviation from the case of only the three yellow DNPs (broken line in Fig.4(d)). Owing to the combined scattered fields by the nanoparticles in the linear chain (waveguide) segment, the electric field spatial profile is slightly altered, and the focal spot electric field enhancement is changed from 5 to 4.6. The location of the focal spot remains unchanged. As the quantum dot excitation is invariably an electric dipole like transition, the enhancement of the excitation can be expressed as the enhancement of the electric field intensity at the location of the QD, expressed here as,

$$\Gamma_{Focus} = |E_x^{(QD)}(\bar{r} = r_{QD})|^2 / |E_{in}(\bar{r} = r_{QD})|^2 \qquad (12)$$

In Fig. 4(e) is shown (dotted line) the spectral spread of the electric field enhancement ($\Gamma_{Focus}$) in the focal spot created by the three DNPs. The peak matches with the in-phase normal mode peak and attains a value above 25. The solid line shows the spectral spread of the field enhancement for the complete structure. The peak position shows slight (<1nm) blueshift and drop in enhancement (22 from 25) owing to the combined effect of the scattering by the additional nanoparticles. This is a minor effect caused by the blue nanoparticles, leaving the focusing function essentially unchanged in the complete multifunctional structure. This is a quite useful property that is exploited for the optimization of the multifunction as it allows tuning the focusing function independent of the lossless propagation.

## B. Emitted Light Propagation

In this subsection we present the analysis underlying the finding of the electric field distribution associated with the 980nm emitted light from the QD embedded in the multifunctional structure shown in Fig.2(c).

In our design of the DNP based multifunctional system the functions of guiding and no-loss propagation of the QD emitted light (at 980nm) are achieved via the collective transverse magnetic dipole ($TE_{1,1}$) and transverse electric dipole ($TM_{1,1}$) modes of the DNP array as depicted in Fig.5(a). The propagating modes of the waveguide are excited by the QD modelled as an ideal transverse electric dipole emitter placed on the axis of the waveguide at a distance D from the surface of the first DNP of the chain. For an axially symmetric situation as depicted in the figure, the transverse source dipole only excites the propagating modes with z angular momentum +1 and -1, resulting in a propagating mode consisting of transverse electric and transverse magnetic modes, as shown in Fig.5(a).

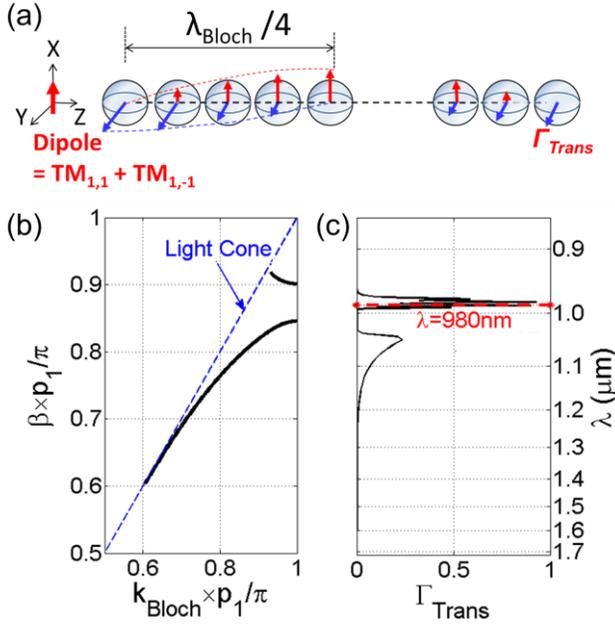

**Figure 5**. (a)Schematic showing the propagation of the transverse dipole mode through a chain of DNPs. Red solid arrows show the transverse electric dipole and blue broken arrows the transverse magnetic dipole mode excitations As shown, both the dipole mode excitations are modulated via the corresponding Bloch wave vector of the propagating mode. (b) Photonic bandstructure for infinite chain of $TiO_2$ BBs with radius 171.3nm and pitch 429nm. The blue dotted line represents the light cone above which the bands become lossy. (c) The energy transfer efficiency of a finite chain of 40 BBs (parameters same as panel (b)), with the source dipole placed D=98.5nm away from the surface of the first BB.

The collective dipole mode excitation can be expressed as a Bloch wave in the periodic array that results in the lossless photonics bands [11]. The solution of these bands are obtained [36] for an infinite chain of $TiO_2$ BBs with radius 171.3nm and pitch 429nm. Figure 5(b) shows the dispersion (photon energy with respect to the Block wave-vector) of the two lossless bands that emerge from the collective oscillation of the transverse electric and transverse magnetic dipole modes in the infinite chain. For a finite chain with N DBBs excited by a transverse electrical dipole [Figure 5(a)], we define the energy transfer factor $\Gamma_{Trans}$ as the magnetic dipole excitation of the last DBB of the array normalized to the source electrical dipole (defined by multipole coefficient $b_{1,\pm1}^{(source)}$) representing the QD, i.e.

$$\Gamma_{Trans} = \frac{\sum_{m=-1}^{1} |a_{1,m}^{(N)}|^2}{\sum_{m=-1}^{1} |b_{1,m}^{(source)}|^2} \qquad (13)$$

It can be thought of as a product of the insertion factor ($\Gamma_{Insert}$) from the dipole to the first DNP of the linear chain and the efficiency of subsequent propagation ($\Gamma_{Propag}$):

$$\Gamma_{Trans} = \Gamma_{Insert} \times \Gamma_{Propag} \qquad (14)$$

The spectrum of $\Gamma_{Trans}$ for a finite linear chain with 40 DBBs with R=171.3nm and $p_1$=429nm, with the source dipole at a distance of D=98.5nm from the surface of the first BB is shown in Figure 5(c). This spectrum closely corresponds to the photonic band structure shown in Figure 5(b). A $\Gamma_{Trans}$ of 0.8 is achieved at 980nm wavelength, which signifies that 80% of the emitted power of the QD is transferred to the magnetic dipole mode of the last DBB. Furthermore, this value is independent of the number of DNPs in the waveguide as the propagation is lossless. The lossless propagation is unaffected as we introduce the two focusing DBB and consider guiding and propagation of the QD emitted photon in the multifunctional structure at 980nm as shown in Figure 6(a). The spectrum of $\Gamma_{Trans}$ for the complete structure is shown with the solid black line in Figure 6(b). The passband and the peak positions are unchanged compared to the case of only the chain (dotted line in Fig. 6(b)), but the value of $\Gamma_{Trans}$ is a little reduced. However, this drop in efficiency is contributed only by the drop in the insertion factor $\Gamma_{Insert}$.

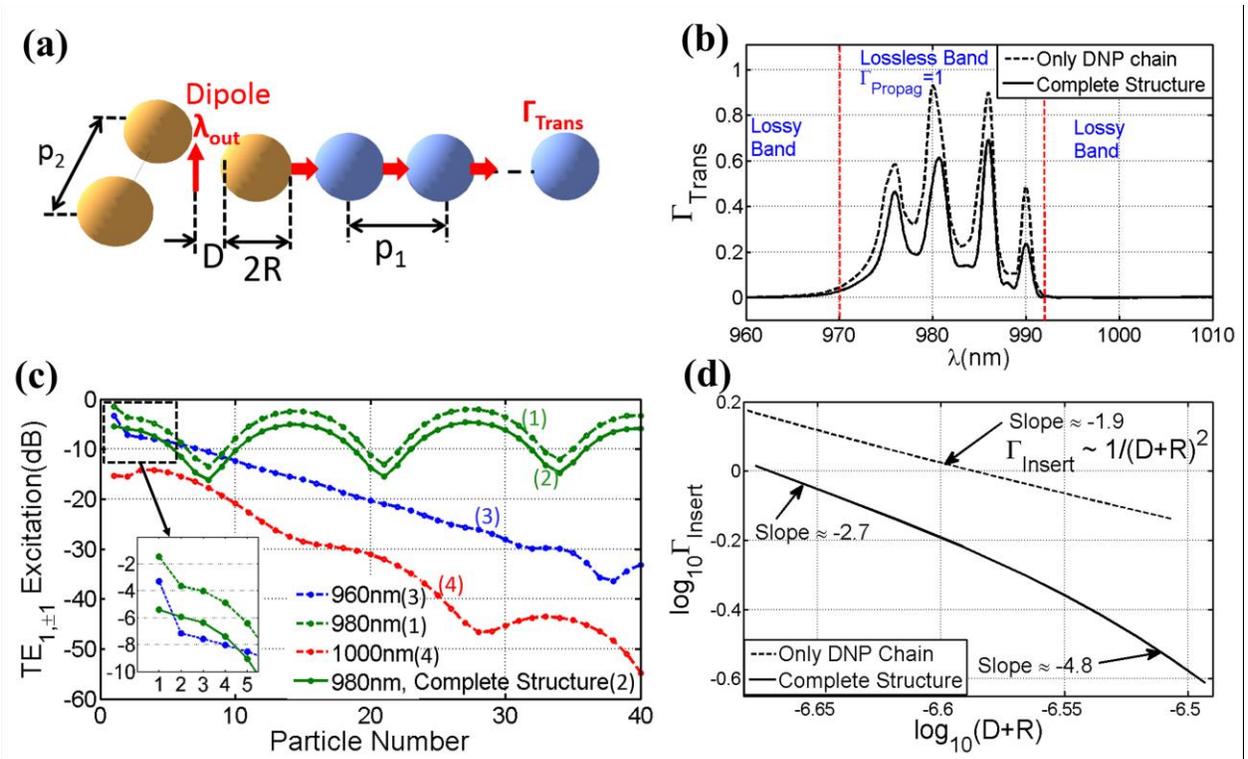

**Figure 6(a)** The complete multifunctional structure for guiding the QD (modeled as transverse electrical dipole) emitted light (red arrows). $\Gamma_{Trans}$ denotes the overall energy transfer efficiency factor. In the calculated results shown, $p_1$=429nm, $p_2$=427.2nm, $R$ = 171.3nm, and $n_i$ = 2.7; **(b)** The dotted line shows the variation of $\Gamma_{Trans}$ as a function of wavelength for a waveguide of 40 DNPs excited by a transverse dipole placed at a distance of 98.5nm from the surface of the first DNP. Equivalent response of the complete structure is shown as the solid line; **(c)** Magnetic dipole mode energy of the DNPs in the array for three different wavelengths signifying the passband and stopbands; **(d)** log-log plot of $\Gamma_{Insert}$ with respect to (D+R) for only chain and for the complete structure.

The lossless propagation is further manifested in Fig.6(c) which shows (green solid line (1)) that the magnetic dipole mode excitations at 980nm remains constant (except for a large period oscillation) as a function of the particle number in the DNP array, while outside the lossless band (blue and red line) drops exponentially. The large period oscillation in the green curve in Fig.6(c) is caused by the interference between the forward propagating wave and the reflection from the abrupt termination of the DNP chain. The multiple backwards and forward reflection from either end of waveguide creates Fabry-Perot modes in the waveguide, which manifest as the almost equal spaced fringes seen in the transfer efficiency function shown in Fig.6(b). In Figure 6(c) the dipole excitation profile for the linear chain only at 980nm is shown using the dotted green curve which makes it clear that the DNP array waveguide propagation still remains lossless in the passband and the decrease in the value of $\Gamma_{Trans}$ is solely caused by the change in $\Gamma_{Insert}$.

The fundamental physical process of insertion is via the coupling of the QD emitter to the propagating mode in the DNP array waveguide. As the QD emitter spontaneously emits a photon, the rate at which the emitted photon is coupled to the DNP array waveguide mode can be derived using the Fermi's golden rule as,

$$\Gamma_{Insert} \propto \frac{2\pi}{\hbar} |D_{if}|^2 |\hat{d} \cdot \bar{E}^{(vac)}_{\bar{k}_{Array}}|^2 \qquad (15)$$

where $D_{if}$ denotes the electric dipole matrix element corresponding to the QD transition with the direction of the dipole denoted by the unit vector $\hat{d}$ and $\bar{E}^{(vac)}_{\bar{k}_{Array}}$ represents the vacuum electric field fluctuation at the location of the QD corresponding to the DNP array propagating mode $|k_{Array}\rangle$. Thus $\Gamma_{Insert}$ also incorporates the anisotropy in the QD emission pattern induced by the DNP structures. Using our formulation based on classical electromagnetism, $\Gamma_{Insert}$ can also be expressed as the interaction between the electric dipole-like emitter with the constituent magnetic dipole mode of the propagating mode of the DNP array. The nearest neighbor interaction between the dipole emitter (transverse electric dipole) and a transverse magnetic dipole mode on the same z axis can be expressed as a function of the center-center distance ($\Delta z$) as,

$$|\tau^{i,i+1}_{1,1,1,1}|^2 \sim \frac{1}{\beta^2 \Delta z^2} |1 + \frac{i}{\beta \Delta z}|^2 \qquad (16)$$

In the dependence of $|\tau|^2$ on $\Delta z$ one can look for the power law dependencies characteristic of the two limiting physical situations for point dipoles, i.e. far field interaction at $\beta \Delta z \gg 1$ with the power law $|\tau|^2 \propto \frac{1}{\Delta z^2}$ (interaction via exchange of real /propagating photons); and near field interaction at $\beta \Delta z \ll 1$ with the power law $|\tau|^2 \propto \frac{1}{\Delta z^\nu}, \nu > 2$ (interaction via exchange of virtual/ evanescent photons). For our case, we thus investigate the variation of $\Gamma_{Trans}$ as D is varied, keeping all the other geometrical parameters the same so that $\Gamma_{Propag}$ does not change in the process. The underlying physics emerges when the dependence of $\Gamma_{Insert}$ is plotted (Fig. 6(d)) on the log-log scale with respect to (D+R), the separation of the dipole emitter to the *center* of the first DNP. For the case of only the simple DNP chain we extract an exponent of ~ -1.9 (dotted line in Fig. 6(d)). That is,

$$\Gamma_{Insert} \sim \frac{1}{(D+R)^2} \qquad (17)$$

implying essentially far-field interaction as the dominant insertion mechanism. However for interaction between distributed collective modes instead of point dipoles, the simple power law signature of point dipoles can no longer be used to distinguish between the near field and far field interaction. This is revealed in the behavior for the complete (i.e. including the yellow DNPs 1 and 2) multifunctional structure plotted as the solid line in Fig. 6(d). Here the QD dipole interacts simultaneously with all three yellow DNPs through their collective mode and the power law is not applicable to help reveal the nature of the photons: virtual or real in the QD-DNP integrated device functions.

**Quantum Dot Spontaneous Emission Rate**
The total rate of spontaneous emission of the QD emitter is obtained by integrating the overlap of the QD transition dipole with the vacuum fluctuation of the electric field corresponding to all the possible $\bar{k}$ states and is given by,

$$\gamma_{Spon} = \frac{2\pi}{\hbar^2} \int_{\bar{k}} \frac{V}{8\pi^3} d\bar{k} \, |D_{if}|^2 |\hat{d} \cdot E_{\bar{k}}^{(vac)}|^2 \, \delta(\omega - ck) \tag{18}$$

Here $V/8\pi^3$ represents the usual normalization factor for the k space integration and $\omega$ denotes the frequency of the emitted photon. We note that the dipole matrix element $D_{if}$ depends on the polarization of the corresponding k states. As the dipole matrix element is factored out from the integral and one gets the familiar spontaneous decay rate in terms of the partial photon local density of states (LDOS) [37],

$$\gamma_{Spon} = \frac{\pi}{3\hbar^2} |D_{if}|^2 \frac{\hbar\omega}{\epsilon} \rho_{\hat{d}}^{\bar{E}}(\omega) \tag{19}$$

where $\rho_{\hat{d}}^{\bar{E}}(\omega)$ is the partial photon LDOS at the location of the QD along the transition dipole direction $\hat{d}$. The permittivity of the medium at the location of the QD is denoted by $\epsilon$. The superscript $\bar{E}$ indicates that only the electric field component of the LDOS [38] is considered since the QD acts like an electric dipole like emitter. We note that the integral of equation (18) computes the total decay rate and hence the information on the anisotropy in the spontaneous emission process leading to the directionality is lost.

In the absence of the DNP assembly, the free space photon LDOS can be expressed as,

$$\rho_o^{\bar{E}}(\omega) = \frac{\omega^2}{2\pi^2 c^3} \tag{20}$$

or,

$$\rho_o^{\bar{E}}(\lambda) = \frac{2\pi}{\lambda^4} \tag{21}$$

In the presence of the DNP structure the photon LDOS at the location of the QD gets modified. If the objective is achieving maximum enhancement in the QD emission rate in the weak coupling regime, known as the Purcell enhancement factor ($F_p$), then maximizing the LDOS at the QD location and at the emission wavelength becomes the guiding criterion for the design. The modified photon LDOS can thus be expressed as,

$$\rho_{\hat{d}}^{\bar{E}}(\lambda) = F_p(\lambda) \frac{2\pi}{\lambda^4} \tag{22}$$

We have calculated the photon LDOS using the well-established Dyadic Greens Function approach [37, 38], where the photon LDOS at any point $\bar{r}$ in space can be expressed as,

$$\rho_{\hat{d}}^{\bar{E}}(\bar{r},\omega) = \frac{3\omega}{\pi c^2} Im[\hat{d} \cdot \vec{G}^E(\bar{r},\bar{r},\omega) \cdot \hat{d}] \tag{23}$$

It is shown in Fig.7 and reveals the change in the spectral distribution of photon LDOS for the DNP based complete structure at the location of the dipole as a function of wavelength showing a Purcell enhancement $F_p \approx 2$. It must be noted that this value of $F_p$ does not reflect any limitation of the DNP based optical systems as no optimization was performed to maximize $F_p$ in this design. It has been demonstrated [19] that Purcell enhancement of the order of hundred is achievable with properly optimized DNP based architectures.

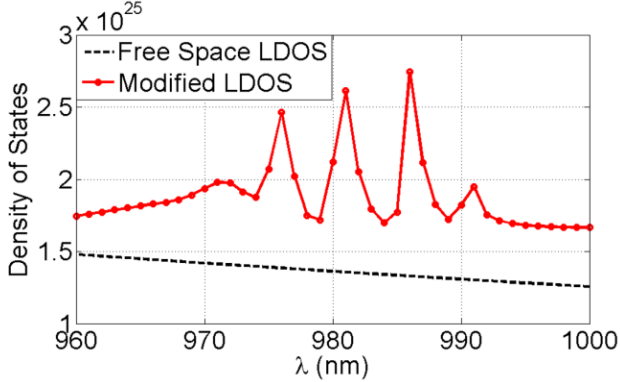

Figure 7. Photon LDOS spectrum showing the free space LDOS (broken curve) along with the modified LDOS by the collective resonance of the DNP assembly.

## IV. SUMMARY AND CONCLUSIONS

In this paper we address, through analytical modeling and numerical simulation, the goal of *on-chip* integration of active and passive optical elements for the realization of nanoscale integrated photonic circuits. The passive components such as a lens and waveguide are implemented using nanoscale (subwavelength) dielectric resonators as building blocks of a single multifunctional structure we denote as LMU (light manipulating unit) as these provide the advantage of significantly reduced footprint with negligible loss and better potential integration with nanoscale light sources such as the nanotemplate-directed epitaxial quantum dots [1, 2, 3]. We show that the needed simultaneous multifunctional behavior at multiple wavelengths in such QD-LMU integrated nanophotonic systems can be achieved by exploiting different collective modes of the interacting magnetic and electric multipole resonances of the DBBs positioned in appropriate geometries that are co-designed with the positioning and working of the active device such as light emitting quantum dots. Specifically, we present here a systematic study of a model multifunctional structure (Fig.2 (a)) made of spherical dielectric nanoparticles that enables an analytic formulation based on multipole expansion method. Other shapes, better suited for optimizing desired functions, can be readily simulated using numerical calculation packages.

As illustration, we demonstrated focusing of incident light at 532nm with a 25 fold enhanced electric field intensity (Fig. 2(b)) at the QD location by using the collective magnetic octupole mode in the co-planar assembly of spherical DNPs. The dominant contribution to this enhancement is from the three yellow DNPs surrounding the QD as seen in Fig. 2(a). Additionally, lossless *on-chip* propagation of the QD emitted light at 980nm is *simultaneously* achieved via excitation of the collective magnetic and electric dipole modes of the DBBs whose electric field distribution is shown in Fig. 2(c). The dominant contribution is from the linear chain part of the entire structure which acts as a waveguide. Quite significantly, our findings reveal that the coupling of the light emitter (such as quantum dot) and the passive element (such as a waveguide) may not be automatically taken to be either near-field or far field in its spatial dependence.

We emphasize that the focusing and propagation wavelengths in the multifunctional structure are very close to the resonance peaks of individual DNPs. Hence, by choosing the material, size, and shape of the constituent DBBs and thereby spectrally tuning the single DBB multipole resonance peaks, the focusing and propagation wavelengths of the LMU can be tuned to desired values. In fact, for the specific architecture (Fig.2) analyzed in this paper, Fig.3(b) indicates that using sub-500nm sized DNPs made of materials such as $TiO_2$ ($n_i \approx 2.7$), GaAs ($n_i \approx 3.5$), and silicon ($n_i \approx 3.8$) the operational wavelength can be tuned over a broad optical spectrum all the way to the near IR communication bands (1300nm and 1550nm) for direct application in the domain of optical communications. We note that fabrication of nanopillar shaped BBs with square or circular base as the constituent building block, readily implemented by nanolithography, enables their monolithic integration with the nanotemplate directed synthesis of regular array of SPSs [1]. For instance, the spectrally uniform InGaAs/GaAs SQD array [1], with planarizing overgrowth of a GaAs layer followed by standard lithographic methods, can be readily integrated with GaAs BB based multifunctional light manipulating units [2, 3]. Such co-designed structures may, for optically pumped emitters, be used for enhancing the electric field at the light wavelength employed for pumping the QD light emitter, or equally aimed at–enhancing the electric field at the emitted photon wavelength (i.e. the Purcell factor) depending upon the objective. Furthermore, it should be noted that the effect of the presence of a Si or GaAs substrate can be explored for its role in affecting the desired functions but we expect that the well-established technologies of lifting-off the QD array as a thin membrane and transferring it to an optically suitable substrate will eliminate the as-grown substrate impact. We conclude noting that the richness of possibilities provides strong incentive to undertake further theoretical and experimental examinations of the on-chip integration of quantum dot emitters with dielectric nanoscale resonator based multi-functional optical systems, co-designed for the smallest foot-print architectures suited for lowest power demand and dissipation coupled with high performance.

## ACKNOWLEDGEMENTS
This work was supported by Army Research Office, Grant No. W911NF-15-1-0298.